\documentstyle[prl,aps,psfig]{revtex}

\begin{document}

\draft

\preprint{}

\def\be{\begin{equation}}

\def\ee{\end{equation}}

\def\bea{\begin{eqnarray}}

\def\eea{\end{eqnarray}}

\title{ Can backscattering off an impurity enhance the current?}

\author{D.E.  Feldman$^{1,2}$ and Yuval Gefen$^1$}

\address{ $^1$Condensed Matter Physics Department, Weizmann Institute of Science,
76100 Rehovot, Israel\\
$^2$Landau Institute for Theoretical Physics, 142432
Chernogolovka, Moscow region, Russia }

\maketitle

\begin{abstract}
It is well known that while forward scattering has no effect on the conductance of one-dimensional systems,
backscattering off a static impurity suppresses the current. We study the effect of a time-dependent
point impurity on the conductance of a one-channel quantum wire. At strong repulsive interaction
(Luttinger liquid parameter $g<1/2$), backscattering renders the conductance greater than its value
$e^2/h$ in the absence of the impurity. A possible experimental realization of our model is a constricted
Hall bar at fractional filling factors $\nu=1/(2n+1)$ with a time-dependent voltage at the constriction.

\end{abstract}

There are several motivations to study one-dimensional (1D) quantum conductors.
Quantum wires are expected to be an essential component of future
nanoelectronic devices \cite{Science}. The analogy between 1D
electron liquid and edge states of the 2D electron gas is
conducive to the understanding of the Quantum Hall effect \cite{QHE}.
There are also many other related systems such as vortex lines in type-II
superconductors \cite{vortex}. From the theoretical point of view
1D conductors are the simplest non-Fermi-liquid systems.
Probably, the most appealing consequence of non-Fermi-liquid behavior is
the existence of fractionally charged quasiparticles,
recently observed in experiments on Quantum Hall systems \cite{fc}.
Since the experimental setup was based on a realization of
1D quantum wire with an impurity, the latter problem 
has received considerable renewed attention. 

The effect of an impurity on a gas of noninteracting electrons is evident.
It leads to backscattering, hence to suppression of the current. 
Qualitatively the same happens in the case of a static impurity
in the presence of electron-electron 
interaction too.
However, in that case the effect turns out to be counterintuitively strong. Even
an arbitrarily weak impurity renders the conductance of a long wire
equal to zero, thus effectively cutting the wire into two independent pieces
\cite{ML,KF,FO}. This is yet another manifestation of strong correlations in a
non-Fermi-liquid state.

While recent activity has focused primarily on the problem
of static impurities, it is also interesting to understand 
what happens if the impurity potential depends on time. This
question touches upon the problem of pumping \cite{pumping}
and the effect of phonon pulses in 1D conductors
\cite{phonons}. Recent works \cite{fl} consider the effect of a
time-dependent impurity on Fermi-liquid states. Much less attention
was devoted to the interplay between a time-dependent impurity
and non-Fermi-liquid effects \cite{Chamon}. In this Letter we study the simplest
question of this type: how a weak,  point-like impurity whose potential
depends
on time affects the conductance of a quantum wire with a
repulsive electron-electron interaction. In the absence of
interaction the answer would be obvious. The impurity would
decrease the current, the suppression of the current 
depending on the strength of the impurity potential.
Surprisingly, for interacting systems the time-dependent
backscattering impurity {\it can enhance}
the conductance. This is the main result of the present Letter.
Such enhancement takes place as the interaction strength exceeds a threshold value.
In terms of the Luttinger liquid parameter $g$ the threshold is $g=1/2$. 


Let us first discuss the origin of the effect qualitatively.
While a purely qualitative discussion is insufficient to explain
the threshold value $g=1/2$, we show below 
that our effect can be derived heuristically from a simple
analysis of the structure of the Hamiltonian. Our detailed
analysis yields quantitative results
concerning the weak impurity limit and is
based on the bosonization technique \cite{bosonization}.

Using the analogy between a quantum wire and edge states
of the 2D electron gas \cite{revWen},
we interpret
backscattering off a weak impurity as 
tunneling between two chiral systems of right- and left-movers.
The tunneling density of states diverges as the energy approaches
the Fermi energy \cite{chiral}. In other words, backscattering
is enhanced in the two following cases:
1) the energy of the incident particle is close to the Fermi energy
of the electrons moving in the same direction;
2) the energy of the backscattered particle is close to the Fermi energy
of the electrons which move in the direction opposite to
that of the incident particle. The left and right Fermi energies
differ by the applied voltage $V$.
These statements about the dependence of the backscattering amplitude on the
energy  do not hold for noninteracting electrons.
In that case one should only note
that scattering 
to an occupied state is impossible. The energy dependence
is more pronounced for the stronger  electron-electron interaction.

Now we are in a position
to consider a time-dependent impurity.
Let us assume for simplicity that the time-dependent potential is harmonic,
 $W(t)\sim U\cos\omega t$.
Note, however, that the current enhancement is possible for a general
periodic potential.
In our qualitative discussion we consider the
case $\hbar\omega>eV;\hbar\omega\approx eV$, where $V$ is the 
voltage and  $e= -|e|$ -- the electron charge.
In this case we predict particularly strong current enhancement.
We assume that the Fermi energy of the particles moving to the left
is $E_{\rm FL}=eV$ and the Fermi energy of the right-movers -- $E_{\rm FR}=0$.
The scattering is inelastic. 
For small $U$ only processes involving an emission (absorption)
of a single quantum are allowed, hence the
energy change is $\pm\hbar\omega$.
There are four backscattering processes (Fig. 1) which we denote as $L^{\pm}, R^{\pm}$,
where the letter shows the type of particles (left- or right-movers)
and the $+/-$ sign corresponds to increasing/decreasing the
energy of the particle. The processes $R^-$ and $L^-$ are suppressed by the
Pauli principle since they lead to the scattering into 
states which are below the Fermi energies, $E_{\rm FL}$ and $E_{\rm FR}$
respectively. 
The processes $L^+$ and $R^+$ drive particles into states
above the Fermi energy. However, only $R^+$ processes lead
to the scattering of the particles with initial energy around $E_{\rm FR}$
to final states with energy close to $E_{\rm FL}$. 
As discussed above, for such particles the probability
of backscattering is enhanced. 
On the other hand, $L(R)^+$ processes
are effective for particles with energies $E$ in the
interval $E_{\rm FL(FR)}>E>E_{\rm FR(FL)}-\hbar\omega$, where
$E_{\rm FL}>E_{\rm FR}$ (cf. Fig. 1).
Hence the number of left-moving particles which potentially
can be subject to strong $L^+$ backscattering processes exceeds the
number of right-movers which potentially are subject to $R^+$ processes \cite{clear}.
As the electron-electron interaction increases,
$R^+$ becomes more important. At some threshold the latter
begins to provide the main contribution to the backscattering current,
hence it determines its direction.
Since this process modifies right-movers to left-movers,
such backscattering enhances the current. Thus, paradoxically,
the scattering of particles backwards adds electrons to the forward flow.

Below we provide a quantitative theory (valid for general $V,\omega$)
and briefly discuss 
a possible experimental realization of our model and the related problem
of a Luttinger liquid affected by an external noise. 

We consider spinless electrons \cite{foot}
and concentrate first on the case of zero temperature.
Our starting point is the Tomonaga-Luttinger model \cite{KF,bosonization}
with a point impurity. The Hamiltonian reads

\bea
\label{1}
H=\int dx [-\hbar v_F(\psi_R^\dagger(x) i\partial_x\psi_R(x)-
\psi_L^\dagger(x) i\partial_x\psi_L(x) ) +
K(\psi^\dagger_R(x)\psi_R(x)+\psi^\dagger_L(x)\psi_L(x))^2
\nonumber \\ 
+\delta(x)W(t)(\psi^\dagger_L(0)\psi_R(0)+\psi^\dagger_R(0)\psi_L(0))],
\eea
where $\psi_R$ and $\psi_L$ are the fermionic field operators of the 
right- and left-moving electrons, $v_F$ is the Fermi velocity,
$W(t)$ the potential of the impurity located at the origin and
$K$ -- the interaction strength. Below we set $v_F$ and $\hbar$
equal to unity. We assume that the right- and left-movers are injected
from the leads with chemical potentials $\mu_R=0;\mu_L=eV>0$.
Note that in our notation $\mu_L$ is located on the
right (Fig. 1).

We follow the technical procedure developed in Ref. \cite{Wen}
for a static impurity. First, we apply the interaction representation
such that the chemical potentials in both leads become zero. This
transformation introduces time dependence into the $\psi_L$
operator and modifies the impurity contribution in Eq. (\ref{1})
which now reads 
$\delta(x)W(t)(\psi^\dagger_L(0)\psi_R(0)\exp(i\omega_0t)+
h.c.)$, where $\omega_0\equiv eV$. Next, we derive an expression for
the current operator.
The current includes the background contribution from the particles
injected from the leads and the backscattering contribution associated with
the impurity. The background contribution is equal
to the current in the absence of the impurity.
Since $\mu_L>0$, the background particle current (of particles with $e<0$)
flows to the left.
The backscattering particle current can be defined \cite{Wen} as
$\hat I_{\rm bs}=d\hat N_L/dt = - d\hat N_R/dt$, where $\hat N_L$ 
and $\hat N_R$ are the particle number operators. A positive value
of the backscattering current corresponds to the enhancement of
the background current. In terms of the $\psi$-fields
the backscattering current operator (in the interaction representation)
is given by the equation

\be
\label{2}
\hat I_{\rm bs}=-iW(t)(\psi^\dagger_L(0)\psi_R(0)\exp(i\omega_0t)-
h.c.).
\ee

Since $W(t)\sim\cos \omega t$, one finds that there
are two types of time-dependent terms in the Hamiltonian 
and the current operator: (i) terms proportional to
$\exp(\pm i(\omega+\omega_0)t)$ and (ii) terms proportional to
$\exp(\pm i(\omega-\omega_0)t)$. If only terms of the first (second)
type existed our problem would be equivalent to a
static impurity in the presence of an external voltage drop
$V_{1(2)}=(\omega_0\pm\omega)/e$. Hence,
the backscattering current can be represented as
$I_{\rm bs}=I_1+I_2+I_{12}$, where $I_{1,2}$ denotes the backscattering
current in the static problem with the voltage $V_{1,2}$, and $I_{12}$
is the 'interference' contribution. We will see that to the lowest
order in the impurity potential the interference current $I_{12}$
is ac, hence the dc-current is made of $I_1$ and $I_2$ only.
Hence if one is interested in the averages over time
$t>1/\omega$, the contribution $I_{12}$ drops out.
We know from Ref. \cite{KF} that $I_{1,2}\sim |V_{1,2}|^{2g-1}$,
where $g=\sqrt{\pi\hbar v_F/(\pi\hbar v_F + 2K)}$ is the $g$-ology parameter
of the Luttinger liquid. For $g<1/2$ the exponent $2g-1$
is negative, hence
the main contribution to
$I_{\rm bs}$ is $I_2$. Thus the direction of the backscattering current
is the same as in the static problem with a voltage $V_2$. 
At $\omega>\omega_0$ the sign of this voltage is opposite to the
sign of the applied voltage $V$. This shows that the backscattering current
{\it enhances} the background one.

In order to actually calculate the currents
$I_{\rm dc}=I_1+I_2$ and $I_{\rm ac}=I_{12}$
we employ the bosonization transformation
\cite{bosonization,Wen}. This leads to the action

\be
\label{3}
L=\int dt dx \biggl[\frac{1}{8\pi}
((\partial_t \hat\Phi)^2-(\partial_x \hat\Phi)^2)
-\delta(x)W(t)(e^{i\sqrt{g}\hat\Phi(t,x=0)}e^{i\omega_0 t} + h.c.)\biggr],
\ee
where $W(t)=U\cos\omega t$ and the bosonic 
 field $\hat\Phi$ is related
to the charge density as $\hat\rho=\sqrt{g}\partial_x\hat\Phi/(2\pi)$.
The same action describes a quantum Hall bar with a 
(time-dependent) constriction \cite{revWen}. In that
case $g$ is the filling factor $g=1/(2n+1)$.
The current operator reads

\be
\label{4}
\hat I_{\rm bs}=-iW(t)e^{i\sqrt{g}\hat\Phi(t,x=0)}e^{i\omega_0 t}+ h.c.
\ee

To find the backscattering current at the moment $t$ we
have to calculate the average

\be
\label{5}
\langle\hat I_{\rm bs}(t) \rangle=
\langle 0|S(-\infty;t) \hat I_{\rm bs} (t) S(t;-\infty)|0\rangle,
\ee
where $\langle 0|$ denotes the ground state, $S$ is the scattering matrix.
To the lowest order in $W$

\be
\label{6}
S(t;-\infty)=1-i\int_{-\infty}^t dt'
[W(t')e^{i\sqrt{g}\hat\Phi(t',x=0)}e^{i\omega_0 t'}+ h.c.];
S(-\infty;t)=S^*(t;-\infty).
\ee

Further calculations follow the standard route \cite{Wen}
and show that the current includes a dc-contribution  $I_{\rm dc}$ and
an ac-contribution $I_{\rm ac}$ of frequency $2\omega$.
We first discuss the more interesting dc-contribution.
It has different forms at 
$\omega>\omega_0=eV$ and $\omega<\omega_0$ (cf. Fig. 2)
\bea
\label{7}
I_{\rm dc}=\frac{U^2}{2(v_F\tau_c)^2}\Gamma(1-2g)\tau_c^{2g}\sin 2\pi g
[(\omega-\omega_0)^{2g-1}-(\omega+\omega_0)^{2g-1}], ~~~
\omega>\omega_0;\\
\label{8}
I_{\rm dc}=-\frac{U^2}{2(v_F\tau_c)^2}\Gamma(1-2g)\tau_c^{2g}\sin 2\pi g 
[(\omega_0-\omega)^{2g-1}+(\omega+\omega_0)^{2g-1}], ~~~
\omega<\omega_0,
\eea
where $\tau_c\sim 1/E_{\rm Fermi}$
is a short-time cutoff.  While at $g>1/2$ the current
$I_{\rm dc}<0$ in both cases, at
strong interaction, $g<1/2$, the backscattering current becomes
positive as $\omega>\omega_0$. In other words,
in the latter case it flows in the direction of the background current.
In the limit $\omega_0\rightarrow 0$ one finds
a correction to the conductance (which would be equal to $G=e^2/h$
in the absence of the impurity \cite{no_g}). The correction
\be
\label{9}
\Delta G = \frac{e^2}{\hbar^3}\left(\frac{U}{v_F\tau_c}\right)^2
(1-2g)\omega^{2g-2}\Gamma(1-2g)\tau_c^{2g}\sin 2\pi g
\ee
is positive for $g<1/2$. The generalization to the case when the time-dependent potential $W(t)$
contains several harmonics is straightforward.

The results (\ref{7},\ref{8}) are obtained at zero temperature.
As $T>|\omega-\omega_0|$ the expression for the current is modified.
The effect of finite temperature can be determined with the Keldysh
technique \cite{keldysh}. At low temperatures
$\omega+\omega_0 > T > |\omega-\omega_0|$ the backscattering current is
given as the sum of two terms, proportional to
$(\omega-\omega_0)T^{2g-2}$ 
and to $(\omega+\omega_0)^{2g-1}$
respectively. The current becomes positive (enhancement of the total current)
for $\omega>\omega_0$,
$T<(\omega+\omega_0)\left(\frac{\omega-\omega_0}{\omega+\omega_0}\right)^{1/(2-2g)}$.
As $T\gg\omega, V$ the dependence
of the backscattering current on $\omega$ drops out. The latter turns
out to be always negative: $I_{\rm bs}\sim - \omega_0 T^{2g-2}$.

There is also an ac contribution of frequency $2\omega$ to the current.
At $T=0$ it reads as follows

\bea
\label{7ac}
I_{\rm ac}(t)=\left(\frac{U}{v_F\tau_c}\right)^2
\Gamma(1-2g)\tau_c^{2g}\cos (\pi g + 2\omega t) \sin \pi g
[(\omega-\omega_0)^{2g-1}-(\omega+\omega_0)^{2g-1}], 
\omega>\omega_0;\\
\label{8ac}
I_{\rm ac}(t)=-\left(\frac{U}{v_F\tau_c}\right)^2
\Gamma(1-2g)\tau_c^{2g}\sin \pi g 
[\cos(2\omega t - \pi g)(\omega_0-\omega)^{2g-1}+
\cos(2\omega t + \pi g)(\omega+\omega_0)^{2g-1}], 
\omega<\omega_0.
\eea

A possible experimental realization of our model is a 
Hall bar with a constriction \cite{webb}. The role of the backscattering
impurity is played by the constriction that gives rise to weak tunneling
of quasiparticles between the two edges. The tunneling amplitude
can then be made time-dependent by application of a time-dependent
gate voltage. If the system is tuned such that it is close
to a resonance \cite{KF} then it can be described by the Lagrangian 
(\ref{3}) with $W(t)=U\cos\omega t$. 
In the absence of tuning a static contribution $W_0$
should be added to $W(t)$. Still if the
inter-edge tunneling is weak and
$\omega$ is greater than the product of the
voltage difference between the edges
and the quasiparticle charge, we
expect an enhancement of the Hall current in the FQHE (at filling factors
$g=1/(2n+1)$) as 
compared with the absence of a time-dependent perturbation
both at $W_0>W$ and $W_0<W$. 
On the other hand, the time-dependent tunneling
decreases the current in IQHE where the filling factor $g=1$.

A more general question concerns the effect of external noise
on the quantum wire. One can consider the tunneling amplitude
$W(t)=\int d\omega [{U'}_\omega \cos\omega t+{U''}_\omega\sin\omega t]$,
where $\langle ({U'}_\omega)^2\rangle=
\langle ({U''}_\omega)^2\rangle=
S(\omega)$ is the spectral function of the noise.
The simplest realization of a random $W(t)$
are thermal fluctuations of the gate
voltage in the set-up discussed in the previous paragraph.
Another related problem is the effect of the irradiation by phonons.
The total backscattering current can be obtained from Eqs. (\ref{7},\ref{8})
integrating over $\omega$ and substituting $2S(\omega)$
for $U^2$. Note that for white noise, $S(\omega)={\rm const}$,
the backscattering current calculated in such way
vanishes at $g<1/2$. Note however, 
that a mathematically ideal white
noise consists of frequencies
which are higher than the Fermi energy. At such frequencies
an approach based on the Luttinger model cannot be used.

In conclusion, we have found that backscattering off a point impurity
can increase the conductance of a quantum wire. This is a manifestation
of the strong electron-electron interaction
as the Luttinger liquid parameter $g<1/2$.

We thank B.L. Altshuler, A.M. Finkelstein, Y. Levinson and Y. Oreg
for useful discussions. 
DEF acknowledges the support by 
the Koshland fellowship and RFBR grant 00-02-17763.
This research was supported by GIF foundation, the US-Israel Bilateral Foundation,
the ISF of the Israel Academy (Center of Excellence), and by the DIP Foundation.

\begin{figure} \label{fig1}
  \hfill
  \psfig{file=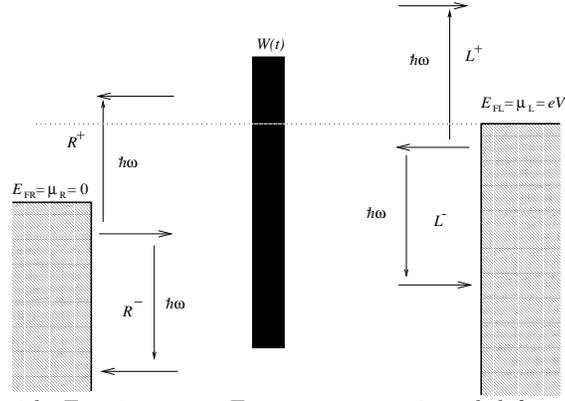,width=2.9in,angle=0}
  \hfill\hfill
\caption{Right-moving electrons with Fermi energy $E_{\rm FR}=\mu_{\rm R}=0$ and left-moving electrons
with Fermi energy $E_{\rm FL}=\mu_{\rm L}=eV$ are backscattered off the time-dependent impurity $W(t)$
via $R^{\pm}$ and $L^{\pm}$ processes in which the electrons gain or lose an energy quantum $\hbar\omega$.}
\end{figure}

\begin{figure} \label{fig2}
  \hfill
  \psfig{file=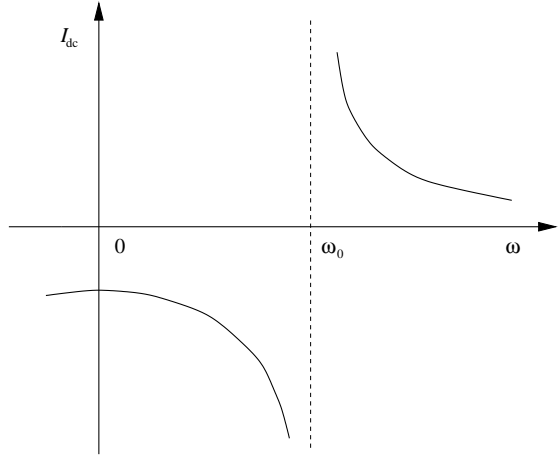,width=2.9in,angle=0}
  \hfill\hfill
\caption{Qualitative dependence of the backscattering contribution $I_{\rm dc}$ to the dc-current
on the frequency $\omega$ at $g<1/2$. $\omega_0=eV/\hbar$, where $V$ is the applied voltage.
Note that $I_{\rm dc}>0$
as $\omega>\omega_0$ (total current enhanced). 
Our approach based on the lowest order of the perturbation theory is insufficient for
the region $\omega\approx\omega_0$.}
\end{figure}


\begin{references}

\bibitem{Science}
See, e.g.,{\it Issues on Nanotechnology}, Science {\bf 290}, 1523 (2000).

\bibitem{QHE}
{\it Perspectives in Quantum Hall Effects}, edited by S. Das Sarma
and A. Pinczuk (John Wiley \& Sons, Inc., New York, 1997).

\bibitem{vortex}
A. Vishwanath and T. Senthil, Phys. Rev. B {\bf 63}, 014506 (2001).

\bibitem{fc}
R. de Piccioto, M. Reznikov, M. Heiblum, V. Umansky, G. Bunin, and D. Mahalu, Nature {\bf 389},
6647 (1997); L. Saminadayar, D.C. Glattli, Y. Jin, and B. Etienne, Phys. Rev. Lett. {\bf 79},
2526 (1997).

\bibitem{ML}
D.C. Mattis and E.H. Lieb, J. Math. Phys. {\bf 6}, 304 (1965).

\bibitem{KF}
C. L. Kane and M. P. A. Fisher, Phys. Rev. B {\bf 46}, 15233 (1992).

\bibitem{FO}
Y. Oreg and A.M. Finkelstein, Phys. Rev. Lett. {\bf 76}, 4230 (1996).

\bibitem{pumping}
B.L. Altshuler and L.I. Glazman, Science {\bf 283}, 1864 (1999).


\bibitem{phonons}
A. J. Kent, D. J. McKitterick, L. J. Challis, P. Hawker,
C. J. Mellor, and M. Henini,
Phys. Rev. Lett. {\bf 69}, 1684 (1992);
V. I. Talyanskii, J. M. Shilton. M. Pepper. C. G. Smith,
C. J. B. Ford, E. H. Linfeld,
D. A. Ritchie, and G. A. C. Jones, Phys. Rev. B {\bf 56}, 15180 (1997).

\bibitem{fl}
See, e.g., S. Datta and M.P. Anantram, Phys. Rev. B {\bf 45}, 13761 (1992);
Y. Levinson and P. Wolfle, Phys. Rev. Lett.  {\bf 83}, 1399 (1999).

\bibitem{Chamon}
P. Sharma and C. Chamon, e-print cond-mat/0011025.


\bibitem{bosonization} 
For reviews see, e.g., S. Rao and D. Sen, e-print cond-mat/0005492;
J. von Delft and H. Schoeler, Annalen Phys. {\bf 7}, 225 (1998).

\bibitem{revWen}
X.-G. Wen, Int. J. Mod. Phys. B {\bf 6}, 1711 (1992).

\bibitem{chiral}
J.M. Kinaret, Y. Meir, N.S. Wingreen, P.A. Lee, and X.-G. Wen, Phys. Rev. B {\bf 46},
4681 (1992).

\bibitem{clear}
Note that when considering the product of the number of particles subject to backscattering
and the probability for such a process, it is not {\it a priori} clear whether it is the $L^+$
or $R^+$ that win.

\bibitem{foot}
As possible realization one may consider a polarized electron liquid or an edge state of 
a quantum Hall bar. For electrons with spin the effective Luttinger parameter $g$ excesses $1/2$
and our effect does not show up.

\bibitem{Wen}
C. de C. Chamon, D. E. Freed, and X. G. Wen, Phys. Rev. B {\bf 51}, 2363 (1995);
throughout the paper we follow the notation of this reference.

\bibitem{no_g}
D.L. Maslov and M. Stone, Phys. Rev. B {\bf 52}, 5539 (1995);
V.V. Ponomarenko, Phys. Rev. B {\bf 52}, 8666 (1995);
I. Safi nad H.J. Schulz, Phys. Rev. B {\bf 52}, 17040 (1995).

\bibitem{keldysh}
J. Rammer and H. Smith, Rev. Mod. Phys. {\bf 58}, 323 (1986).

\bibitem{webb}
F. P. Milliken, C. P. Umbach, and R. A. Webb, Solid State Commun. {\bf 97},
309 (1996).

\end{references}
\end{document}